\documentclass[12pt,preprint]{aastex}

\newcommand{\kms}{km\,s$^{-1}$}

\shorttitle{Core collapse explosion for SN2002ic}
\shortauthors{Benetti et al.}

\begin{document}


\setcounter{footnote}{-1}

\title{Supernova 2002ic: the collapse of a stripped-envelope, massive 
star in a dense medium ?}


\author{S. Benetti\altaffilmark{1}; E. Cappellaro\altaffilmark{1};
M. Turatto\altaffilmark{1}; S. Taubenberger\altaffilmark{2};
A. Harutyunyan\altaffilmark{1}; S. Valenti\altaffilmark{3,4} }

\altaffiltext{1}{INAF-Osservatorio Astronomico, vicolo
dell'Osservatorio 5, 35122 Padova, Italy; stefano.benetti@oapd.inaf.it, enrico.cappellaro@oapd.inaf.it, massimo.turatto@oapd.inaf.it, avet.harutyunyan@oapd.inaf.it}

\altaffiltext{2}{Max-Planck-Institut f\"ur Astrophysik,
Karl-Schwarzschild-Str. 1, 85741 Garching bei M\"unchen, Germany; tauben@MPA-Garching.MPG.DE}

\altaffiltext{3}{European Southern Observatory,
Karl-Schwarzschild-Str. 2, 85748 Garching bei M\"unchen, Germany; svalenti@eso.org}

\altaffiltext{4}{Dipartimento di Fisica, Universit\'a di Ferrara, via
Paradiso 12, 44100 Ferrara, Italy}

%




\begin{abstract}

We revisit the case of SN2002ic that recently revived the debate about the progenitors of
SNeIa after the claim of the unprecedented presence of hydrogen lines over a diluted SNIa
spectrum. As an alternative to the previous interpretation, we suggest that SN2002ic actually
was a type Ic SN, the core collapse of a massive star which lost its hydrogen and helium
envelope. In this scenario the observed interaction with a dense circumstellar material
(CSM) is the predictable consequence of the intense mass-loss of the progenitor and/or of
the presence of a gas rich environment. With this view we establish a link between energetic
SNeIc and highly interacting SNeIIn and adds some credits to the proposed association of
some SNeIIn to GRBs.

\end{abstract}

\keywords{supernovae: SN2002ic, general}

\section{Introduction}\label{intro}

Supernovae, the catastrophic events terminating the life of several kinds of stars, arise
from two fundamentally different paths \citep{tur03}: the core collapse of a massive
star which has exhausted the nuclear fuel or from the thermonuclear explosion of a white
dwarf that overcomes the Chandrasekhar mass limit (about 1.4M$_\odot$) after accretion of
material from a companion star. 

In the favourite scenario for SNeIa progenitors, called single-degenerate (SD), the
companion is a star in a late stage of evolution donating mainly hydrogen and helium. A
viable alternative is that the companion is also a degenerate star (DD scenario), from which
tidal forces strip CO-rich material \citep{hil00}. It is a major concern for the use of
SNeIa as cosmological probes that we do not yet have direct evidence of the nature of their
progenitor systems.

The recent observations of SN2002ic, an alleged type Ia SN showing the unambiguous signature
of hydrogen lines in the spectrum, were claimed to mark a point in favour of the SD scenario
\citep{ham03}. The unusual spectrum of this SN has been explained by the
composition of the spectrum of a typical SNIa plus a contribution due to interaction of the
outer ejecta with a dense CSM. The occurrence of interaction is also seen in the complex
profile of the H$\alpha$ emission, a narrow component on top of a very broad base
\citep{ham03}. With time, the contribution due to interaction becomes dominant, the underlying SNIa
spectrum is washed out, and eventually, two months after the discovery, the spectrum of
SN2002ic closely resembles those of the strongly interacting SNeIIn 1997cy
\citep{ger00,tur00} and 1999E \citep{rig03}.

\begin{figure}
\figurenum{1}
\includegraphics[width=0.4\textwidth]{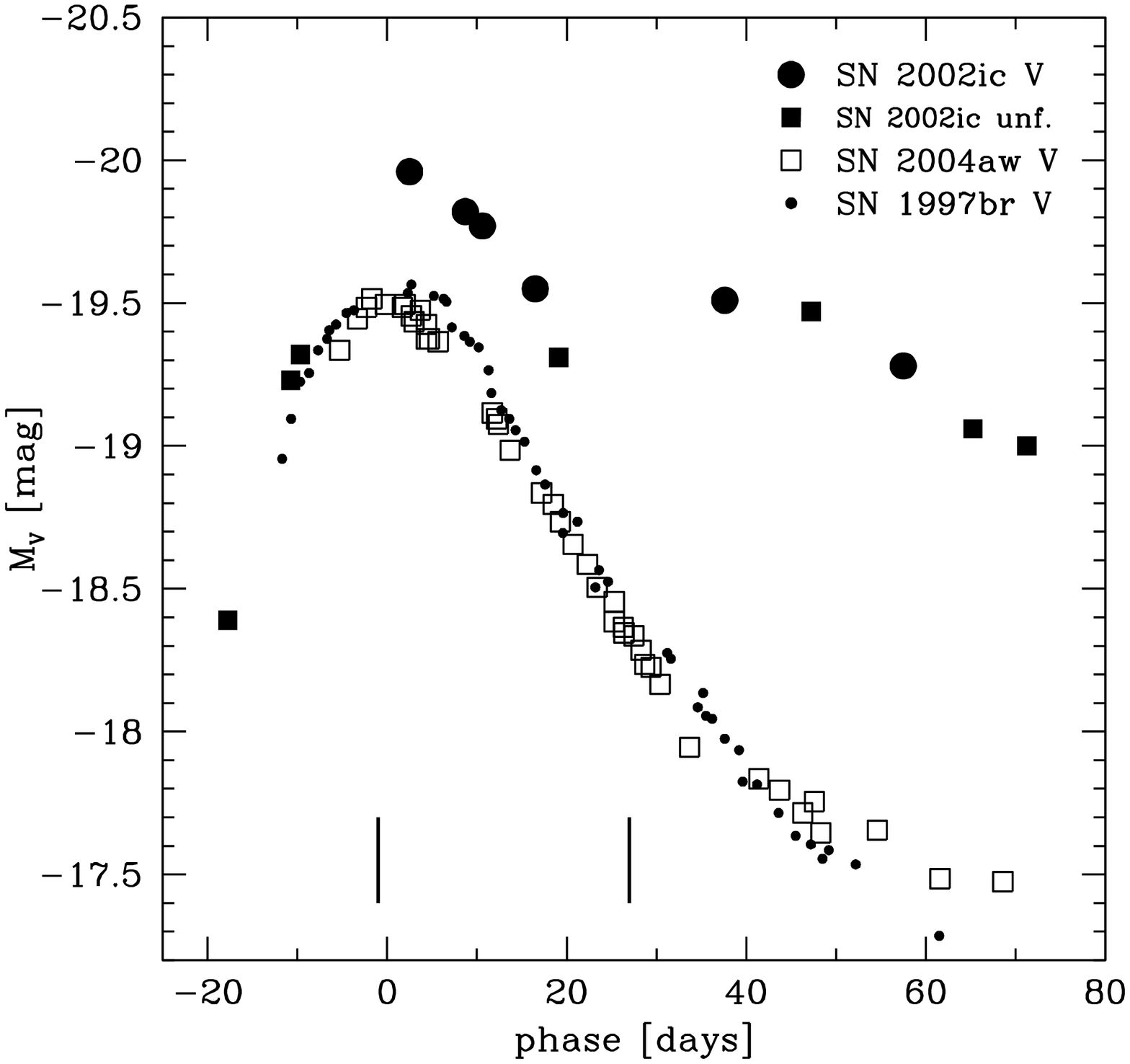}
\caption{Comparison of the V absolute light curves of SNe 2002ic \citep{ham03, woo04}, 2004aw \citep{tau06} and 1997br \citep{li99,alt04}. The SN2002ic points have been K-corrected. Reddening and distance adopted for the three SNe are
reported in the caption of Fig.2. The epochs of the two SN2002ic
spectra shown in Figures 2 and 3 are marked with vertical lines.}\label{lc}
\end{figure}

The light curve of SN2002ic (Fig.\ref{lc}) appears consistent with this interpretation
\citep{woo04,woo06}: after a peak brighter than that of the brightest, slowly declining
SNeIa\footnote{The rise to maximum of SN2002ic seems to take over 20d, which is a few days
longer than in a typical SNIa (or even SNIb/c). Note, however, that the early magnitudes
(square symbols in Fig.\ref{lc}) are highly uncertain.} it begins the usual rapid decline
which, however, is interrupted by a re-brightening one month after maximum. After that, the
late time decline was much slower than typical for SNeIa. This luminosity
evolution, showing a delay between the explosion and the onset of the strong ejecta-CSM
interaction suggests the presence of a cavity in the CSM around the progenitor
\citep{woo04,woo06}.

The detection of strong hydrogen lines in the spectrum of a thermonuclear SN has stimulated
the speculations on the possible nature of the progenitor system. Among the proposed
scenarios are the explosion of a WD in a binary system with a post-AGB companion
\citep{ham03}, the explosion of a C/O core of a 25M$_{\odot}$ star \citep[SN1.5,][]
{ham03,ims05}, the merger of a WD with the core of an AGB star \citep{liv03} and the
explosion of a WD in a supersoft X-ray system \citep{han06}. In general, it is agreed that
the discovery of hydrogen lines favours the SD scenario although the rareness of events like
SN2002ic casts some doubts that they are really representative of all SNeIa \citep{liv03}.
So far only one more SN, SN2005gj \citep{pri05}, has been found to show many similarities
with SN2002ic \citep{ald06}. In that case however, the features which were claimed to
identify SN2002ic as a type Ia were less evident and the first featureless
SN2005gj spectrum is reminiscent of those arising from the hot photospheres of CC
explosions.


\section{Was SN2002ic really a type Ia SN?}\label{sn02ic_noint}

Despite the deep implications of the observations of SN2002ic, very little debate has taken
place on the robustness of the proposed interpretation. It is true that the main features in
the SN2002ic spectrum are fairly well reproduced by the composition of SN1999ee spectrum,
with its characteristic SiII and SII lines, and a (somehow arbitrary) continuum, but,
at a closer look, there appear also some annoying discrepancies. In particular, the spectrum of
SN2002ic obtained in proximity to the maximum light does not show the strong H \& K CaII feature at
$\sim3700$\AA\/ typical of SNeIa \citep[top panel of Fig. 3 of][]{ham03}. About one month later (panel b) of Fig. 3 of Hamuy et al.) a major
discrepancy is seen around 6500\AA: the broad emission attributed to SiII and FeII in SNeIa
is much stronger than in the spectrum of SN2002ic, where, for a direct comparison, one
should also subtract the H$\alpha$ contribution.


Starting from these inconsistencies, we have explored if other interpretations are viable. 

The main argument for the classification of SN2002ic as type Ia is the presence of the
putative Si feature at 6150\AA. Yet other SN types, in particular SNeIc, usually associated
to the collapse of bare CO cores, show a similar feature although the ion identification may
be different \citep[$H_{\alpha}$,][]{bra06}. A recent, intriguing example was that of SN2004aw \citep{tau06}
which indeed on the basis of a single spectrum was initially (mis-)classified as  slow
decliner SNIa, similar to SN1991T \citep{ben04}. The error was corrected after considering
the spectral evolution \citep{fil04}, and eventually the nebular spectrum, dominated by O
and Ca emissions, definitely solved the ambiguity.
However, it remains that the spectra of some SNeIc at early phases can easily be confused
with those of slow decliner SNeIa. 

Contrary to SNeIa, SNeIc show a wide range in their peak luminosities and intrinsic colors,
and the reddening is hard to estimate. In the case of SN2004aw, \cite{tau06} adopted a color
excess E(B-V)=0.37 based on a conservative relation between reddening and the equivalent
width (EW) of the interstellar NaID lines. The adoption of a steeper relation \citep{tbc}
allows for a value as high as E(B-V)$\simeq 0.85$, which has also the benefit of a better
match of the color curves of SN2004aw with those of other objects of the same type.
Therefore, in the following, spectra and magnitudes of SN2004aw have been corrected using
this latter value. With this choice the absolute magnitude at maximum of SN2004aw is even
brighter than that of the hypernova SN1998bw. We stress that the conclusion of our paper
would not change significantly if we had adopted a different value for the reddening.

As an objective test we have applied the automatic classification
software developed by our group \citep{avi05} to the spectra of SN2002ic.
The spectra of SN2002ic have been compared to those
present in the whole Asiago-Padova Supernova archive, which contains about 2700 spectra of
about 380 SNe of all types. The best match is found with the spectra of SN2004aw, 
followed by those of SN1997br, a slow declining SNIa similar to SN1991T.

\begin{figure}
\figurenum{2}
\includegraphics[width=0.5\textwidth]{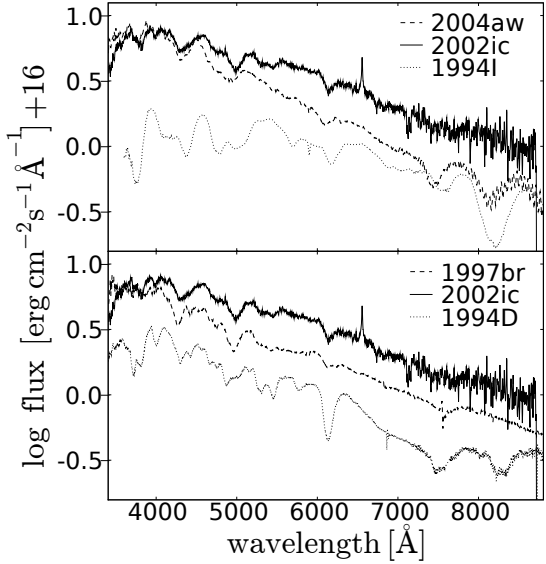}
\caption{Comparison of maximum-light spectra of SN2002ic with those of 
the type Ic SNe 2004aw and 1994I (from Asiago-Padova supernova and McDonald archives) (top panel) and the type Ia SNe 1997br (from Asiago-Padova archive) and 1994D \citep[from][]{pat96} (bottom panel).
The spectra of SNe 2002ic, 2004aw and 1997br are in the parent galaxy rest frame, reddening-corrected and scaled in flux to the distance of SN2002ic
assuming the following values: SN2002ic (2002/11/29): v$_{hel}$ = 19800
\kms, E(B-V)=0.073, $\mu =37.33$; SN2004aw (2004/03/24): v$_{hel}$= 4900
\kms, E(B-V)=0.85, $\mu =34.17$; SN1997br (1997/04/16): v$_{hel}$ = 2069
\kms, E(B-V)=0.35, $\mu =32.25$.
The spectra of SN1994I and SN1994D, corrected as above, are displayed as prototypes of the two classes (SN1994D (1994/03/19): v$_{hel}$ = 450 \kms, E(B-V)=0.00; SN1994I (1994/04/09): v$_{hel}$ = 461 \kms, E(B-V)=0.30).}\label{max_spec}
\end{figure}

Fig.~\ref{max_spec} shows the comparison at maximum light.
In the top panel SN2002ic is compared with SNeIc, 
SN2004aw and SN1994I. 
The match with SN2002aw is good concerning both overall appearance and flux intensity.
The SiII line (or H$\alpha$) is in
place as well as the two absorptions in correspondence to the SII lines although they are much
shallower in SN2004aw than in SN2002ic.
SN1994I, instead, is definitely redder and shows more developed bands not entirely
matching those of SN2002ic. This not surprising considering the broad range of properties of 
SNeIc.

The comparison of SN2002ic with SNeIa is also very interesting (Fig.~\ref{max_spec} 
bottom panel). SN1997br, whose spectrum is dominated by FeII and FeIII lines, 
produces a generally good match: the SII lines at $\sim 5500$ \AA\/ and the
H \& K CaII absorption are weaker than in normal SNIa. Instead, the match with the spectra
of the typical SN1994D is definitely not as good.

SN2004aw spectrum is certainly a better match below $\sim4500$\AA, while SN1997br
better reproduces the featureless continuum above $\sim7000$\AA. 
The fact that both objects produce satisfactory fits to the spectrum of SN2002ic
is not surprising in the light of the misclassification of SN2004aw mentioned above.

\begin{figure}
\figurenum{3}
\includegraphics[width=0.5\textwidth]{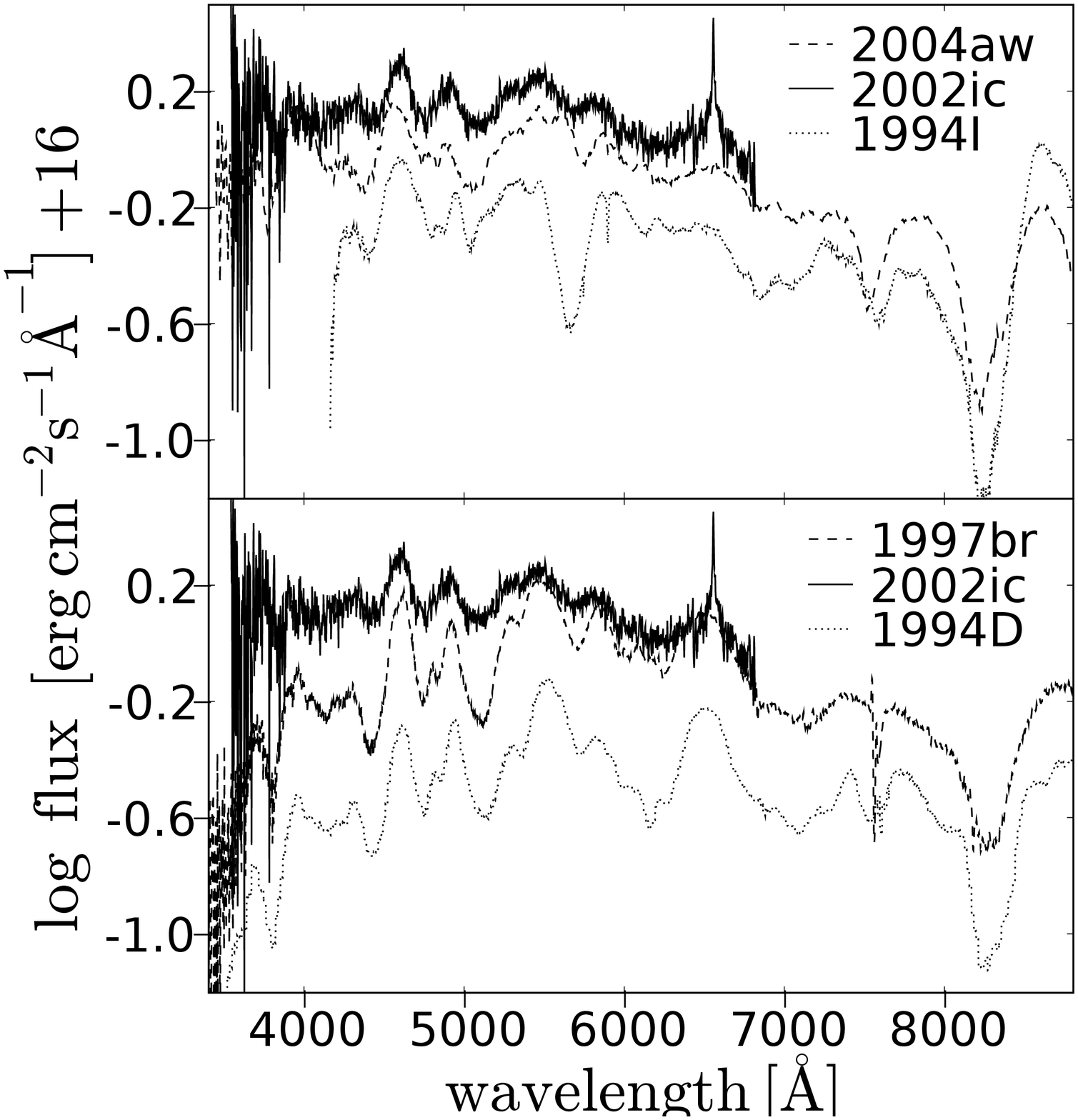}
\caption{Comparison of the 1-month spectrum of SN2002ic (2002/12/27) with those of the type Ic SNe 2004aw (2004/04/20) and 1994I (1994/05/03) (top panel), and the type Ia SNe 1997br (1997/05/13) and 1994D (1994/04/14) (bottom panel). See Fig.~2 for additional information.}\label{month_spec}
\end{figure}

The comparison of the spectra taken one month later, shown in Fig. ~\ref{month_spec}, is
more convincing. The SED of SN2004aw well matches the spectrum of SN2002ic with no
need of any continuum. Some differences can be seen in the intensity of the CaII line at
$\sim 4000$\AA\/ and in the NaID line at $\sim 5700$\AA\/, both stronger in SN2004aw than in
SN2002ic, but the coincidence of strength and position of all other features is good.
We note, in particular, that the broad asymmetric emission at the base of the narrow
H$\alpha$ of SN2002ic is present also in 2004aw and that a weak, narrow H$\alpha$ emission
seems to remain in some SN2004aw spectra. 
 At this phase the spectrum of
SN1994I shows a huge NaID-HeI5876\AA~ feature seen neither in SN2004aw nor in SN2002ic.
SNe 1997br and 1994D are quite similar to each other, as usually do SNIa
when deeper layers are exposed (bottom panel
of Fig. 3) and, both make a reasonable fit to SN2002ic. 
However, the centre wavelengths of the broad emission at $\sim 6500$\AA\/ (SiII,
FeII) is bluer by $\sim 50$ \AA\/ in the SNIa. Moreover, as
already mentioned, the intensity of their 6500 \AA~ emission is much stronger
compared to that of SN2002ic. In general, to have a better match at this phase, the
SNIa spectra have to be "diluted" by a continuum possibly arising from the interaction between the ejecta
and the CSM as suggested in \citet{ham03}.

We may note that the comparison of the light curves shown in Fig.~\ref{lc} is inconclusive
for the identification of the nature of SN2002ic. In fact the absolute V light curves of the
type Ia SN1997br and type Ic SN2004aw are almost identical until almost two months from
explosion. The light curve of SN2002ic also has a similar shape, at least until 20 days
after maximum, but appears 0.5 mag brighter.

\section{The ejecta-CSM interaction in SN2002ic}\label{sn02ic_int}

The re-brightening of SN2002ic, occurring at about one month after maximum and attributed to
the onset of a strong ejecta-CSM interaction \citep{ham03}, averts the light curve from the
normal decline of SNeIa and Ic (Fig.\ref{lc}). Indeed, the spectrum of SN2002ic of Jan.9,
2003, is dominated by a strong H$\alpha$ emission with broad wings and a blue continuum, and
resembles that of a strongly interacting SNIIn.

Although at this point it becomes difficult to disentangle the intrinsic features of the SN
from those of the CSM interaction, a comparative analysis of the late time spectra continues
to support the idea that SN2002ic was a CC SN of type Ic.
In Fig. \ref{nebular} the spectrum of SN2002ic obtained 10 months after the explosion is compared with spectra of the strongly interacting SN1999E, the type Ic hypernova SN1998bw, SN2004aw and the type Ia SN1991T. Also for this spectrum we have run the automatic classification software. Once the H$\alpha$ emission is masked, the best match is with SN1998bw (SNe 1997cy and 1999E spectra excluded). 
While Ca lines are common to all SN types, we believe that the  presence of Mg and O emissions, although not as strong as in normal CC SNe (probably because of the different physical conditions, higher density, due to interaction) are an indication that considerable amounts of intermediate mass elements are present in the shocked ejecta and CSM of SN2002ic, thus supporting the massive progenitor scenario. We have to remind that, on the other hand, \citet{chugai} have modelled the late spectrum of SN2002ic with a mixture of iron group elements plus Ca, reproducing the quasi-continuum and the CaII features with the noticeable exception of the MgI] at 4600\AA.
Note that the oxygen expansion velocity deduced from the [OI] $\lambda \lambda 6300, 6364$ emissions in SNe 2004aw and 1998bw ($\sim 7200$ \kms) is comparable to that of SN2002ic \citep{den04}. Spectropolarimetry of this latter SN has shown that the hydrogen line emitting region is asymmetric \citep{wan04}, which may be due to asymmetries either of the explosion or of the CSM, similarly to SN1999E, for which the asymmetric collapse of a CO core was proposed \citep{fil00}.

\section{Discussion}

 Although the SNIa scenario proposed by \citet{ham03} cannot be ruled out by our analysis, we do believe that there are evidences in favour of the association of SN2002ic to a type Ic SN surrounded by a structured H-rich CSM, possibly asymmetric.

At early epochs SN2002ic interacts only weakly with the fast wind of the progenitor star, as observed also for other SNeIb/c in radio and X-rays \citep{fil00}. After one month the interaction reinforces, likely because the ejecta reach a circumstellar shell of higher density. From that moment on, the light curve of SN2002ic \citep{woo04} closely resembles that of SN1997cy, for the which modelling requires a few solar masses of H-rich material \citep{tur00} in the CSM. The dense CSM is likely the relic of a powerful stellar wind which was active until shortly before the explosion. It has recently been realized that the He cores of massive, rapidly rotating stars can considerably increase in mass. In such cases violent explosions can take place at the ignition of oxygen burning, leading to the ejection of some solar masses of surface material (probably He-rich) months or years before the SN explosion \citep{woos06}. Observational evidences that some massive stars undergo violent mass-loss episodes soon before explosion are also available \citep{ben99}.

Alternatively, if the progenitor was in a binary system, the dense CSM might originate from the companion. Indeed, in the solar neighbourhood a large fraction of WR stars, which are among the best candidates for SNIc progenitors, are in binary systems with hot companions \citep{mof86}. Therefore, contrary to the low-mass, evolved-progenitor scenario, in the case of a young, massive progenitor the presence of a dense CSM is certainly not unexpected.

Our interpretation of SN2002ic with a CC SN would remove this event from the list of arguments concerning the true nature of SNIa progenitors, for which direct information would remain absent. 

\begin{figure}
\figurenum{4}
\includegraphics[height=10cm, angle=0]{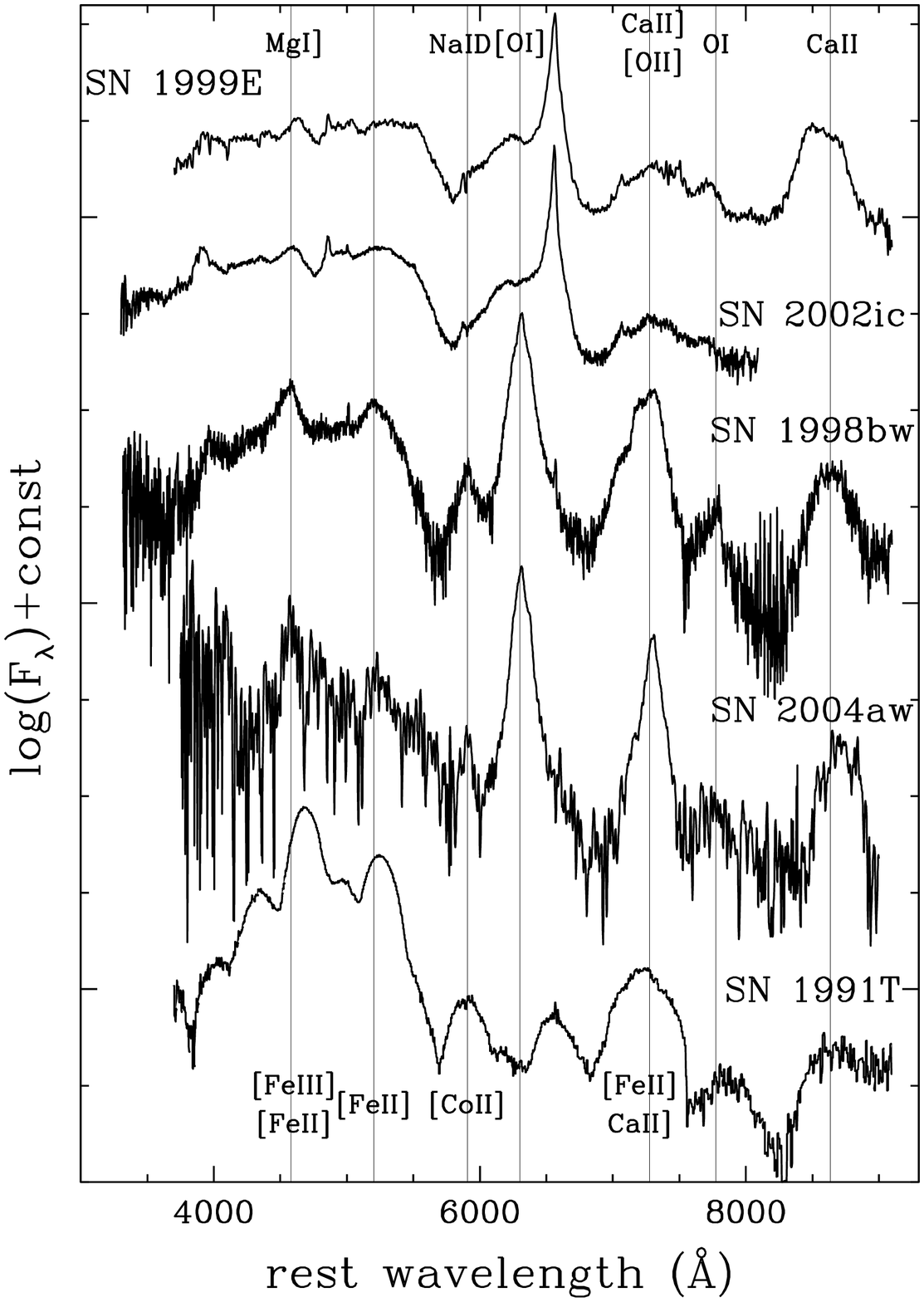}
\caption{Comparison of late time nebular spectrum of SN2002ic \citep[$\phi =+241$d;][]{wan04} with those of SN1999E \citep[$\phi =+237$d;][]{rig03} (IIn), SN1998bw \citep[$\phi =+200$d;][]{pat01} (Ic peculiar), SN2004aw \citep[$\phi =+236$d;][]{tau06} (Ic) and SN1991T \citep[$\phi =+283$d;][]{tur96} (slow decliner Ia). The spectra of the SNe 1999E, 1998bw and 1991T have been reported to the parent-galaxy rest frame and reddening corrected as follows: SN1999E: v$_{hel}$=7800 \kms, E(B-V)=0.27; SN1998bw: v$_{hel}$=2550 \kms, E(B-V)=0.06; SN1991T: v$_{hel}$=1736 \kms, E(B-V)=0.14 \citep{phi99}. For others see caption of Fig.2. The most significant features of SNeIc are indicated by vertical lines and labelled on top of the figure. Bottom labels indicate the position of the most important nebular SNIa features.}\label{nebular}

\end{figure}

On the other side, the metamorphosis of SN2002ic would establish a link between energetic SNeIc (SN2004aw) and highly interacting type IIn (SNe 1997cy, 1999E), which, in turn, would allow a new look to the proposed association of the latter SNe with two BATSE GRBs \citep{ger00,tur00,rig03}. Indeed, although there is no information on a possible GRB associated to SN2004aw, the association of some long GRBs with energetic SNIc is now firmly established \citep{woos06}. In this view, the remarkable similarity of the Jan.9 spectrum of SN2002ic with the first available spectrum of SN1997cy suggests that the two spectra have been obtained at a similar phase and is fully consistent with the epoch of explosion for SN1997cy inferred from its proposed association with the GRB970514. In the same respect, also the spectral analogies found at late times with SN1998bw (Fig.\ref{nebular}) are interesting.

Little credit was given to the association of some SNeIIn with GRBs \citep{woos06}, mainly based on the theoretical argument that a large H envelope was expected to quench the relativistic jet which is supposed to originate the GRB. The presence of a cavity around the exploding star, as seen from the observations of SN2002ic, may leave room for the jet to extend before it shocks into the CSM. SN2001ke, associated to GRB011121, well fits in this scenario. Indeed, \citet{gar03} found that the SED of this event about one month after the burst is similar to that of the highly interacting SNIIn 1998S. Moreover the overall appearance of the SN2001ke spectrum is also reminiscent of those of SNe 2002ic and 2004aw taken at similar phases.

If these were common events, there should be a number of cases where the GRB optical afterglow would remain bright for a longer time than usually observed. Because this is not the case, we do not expect that strongly interacting type IIn SNe are associated to a major fraction of long GRBs.

\acknowledgments 

We are indebted to M. Hamuy and L. Wang for providing the data of SN2002ic. S.B. also acknowledge K. Nomoto for helpful discussions. This work was supported by grants n. 2004029938 of the Italian MIUR and HPRN-CT-2002-00303 of EU-RTN. This work is partially based on observations collected 
at the ESO, Chile (ESO Programme 59.D-0332)

\clearpage

\end{document}